\documentclass[aps,prl,twocolumn,tightenlines,superscriptaddress,showpacs,preprintnumbers,amsmath,amssymb]{revtex4}

\usepackage{graphicx}
\usepackage{dcolumn}
\usepackage{bm}
\usepackage{pstricks}
\usepackage{pst-node}

\graphicspath{{ps}}

\usepackage{subfigure}
\usepackage{mathrsfs}
\usepackage{amssymb}
\usepackage{amsmath}
\usepackage{tabularx}
\usepackage{rotating}
\usepackage{multirow}

\def\dE{{\Delta E}}

\def\mbc{{M_{\rm bc}}}
\def\md{{$B^{+} \rightarrow D^{*+} \pi^0$}}
\def\dstrho{{$\overline{B}{}^{0} \rightarrow D^{*+} \rho^-$}}
\def\GeV{\,{\rm GeV}}
\def\GeVc{\,{\rm GeV}/c}
\def\GeVcc{\,{\rm GeV}/c^2}

\def\MeVcc{\,{\rm MeV}/c^2}

\begin{document}

\preprint{\vbox{ \hbox{   }
                        \hbox{Belle Preprint 2008-09}
                        \hbox{KEK Preprint 2008-02}
}}

\title{ \quad\\[0.5cm]
\boldmath Search for $B^+ \to D^{*+} \pi^0$ decay }

\affiliation{Budker Institute of Nuclear Physics, Novosibirsk}
\affiliation{Chiba University, Chiba}
\affiliation{University of Cincinnati, Cincinnati, Ohio 45221}
\affiliation{The Graduate University for Advanced Studies, Hayama}
\affiliation{Hanyang University, Seoul}
\affiliation{University of Hawaii, Honolulu, Hawaii 96822}
\affiliation{High Energy Accelerator Research Organization (KEK), Tsukuba}
\affiliation{Institute of High Energy Physics, Chinese Academy of Sciences, Beijing}
\affiliation{Institute of High Energy Physics, Vienna}
\affiliation{Institute of High Energy Physics, Protvino}
\affiliation{Institute for Theoretical and Experimental Physics, Moscow}
\affiliation{J. Stefan Institute, Ljubljana}
\affiliation{Kanagawa University, Yokohama}
\affiliation{Korea University, Seoul}
\affiliation{Kyungpook National University, Taegu}
\affiliation{\'Ecole Polytechnique F\'ed\'erale de Lausanne (EPFL), Lausanne}
\affiliation{Faculty of Mathematics and Physics, University of Ljubljana, Ljubljana}
\affiliation{University of Maribor, Maribor}
\affiliation{University of Melbourne, School of Physics, Victoria 3010}
\affiliation{Nagoya University, Nagoya}
\affiliation{Nara Women's University, Nara}
\affiliation{National Central University, Chung-li}
\affiliation{National United University, Miao Li}
\affiliation{Department of Physics, National Taiwan University, Taipei}
\affiliation{H. Niewodniczanski Institute of Nuclear Physics, Krakow}
\affiliation{Nippon Dental University, Niigata}
\affiliation{Niigata University, Niigata}
\affiliation{University of Nova Gorica, Nova Gorica}
\affiliation{Osaka City University, Osaka}
\affiliation{Osaka University, Osaka}
\affiliation{Panjab University, Chandigarh}
\affiliation{Saga University, Saga}
\affiliation{University of Science and Technology of China, Hefei}
\affiliation{Seoul National University, Seoul}
\affiliation{Sungkyunkwan University, Suwon}
\affiliation{University of Sydney, Sydney, New South Wales}
\affiliation{Toho University, Funabashi}
\affiliation{Tohoku Gakuin University, Tagajo}
\affiliation{Tohoku University, Sendai}
\affiliation{Department of Physics, University of Tokyo, Tokyo}
\affiliation{Tokyo Institute of Technology, Tokyo}
\affiliation{Tokyo Metropolitan University, Tokyo}
\affiliation{Tokyo University of Agriculture and Technology, Tokyo}
\affiliation{Virginia Polytechnic Institute and State University, Blacksburg, Virginia 24061}
\affiliation{Yonsei University, Seoul}
  \author{M.~Iwabuchi}\affiliation{The Graduate University for Advanced Studies, Hayama} 
  \author{M.~Nakao}\affiliation{High Energy Accelerator Research Organization (KEK), Tsukuba} 
  \author{I.~Adachi}\affiliation{High Energy Accelerator Research Organization (KEK), Tsukuba} 
  \author{K.~Arinstein}\affiliation{Budker Institute of Nuclear Physics, Novosibirsk} 
  \author{V.~Aulchenko}\affiliation{Budker Institute of Nuclear Physics, Novosibirsk} 
  \author{T.~Aushev}\affiliation{\'Ecole Polytechnique F\'ed\'erale de Lausanne (EPFL), Lausanne}\affiliation{Institute for Theoretical and Experimental Physics, Moscow} 
  \author{A.~M.~Bakich}\affiliation{University of Sydney, Sydney, New South Wales} 
  \author{V.~Balagura}\affiliation{Institute for Theoretical and Experimental Physics, Moscow} 
  \author{E.~Barberio}\affiliation{University of Melbourne, School of Physics, Victoria 3010} 
  \author{A.~Bay}\affiliation{\'Ecole Polytechnique F\'ed\'erale de Lausanne (EPFL), Lausanne} 
% ...
\collaboration{The Belle Collaboration} 
\noaffiliation

\begin{abstract}
We report on a search for the doubly Cabibbo suppressed decay 
$B^+ \to D^{*+} \pi^0$, based on a data sample of 
$657 \times 10^6\, B\overline{B}$ pairs collected at the
$\Upsilon(4S)$ resonance with the Belle detector at the KEKB
asymmetric-energy $e^+ e^-$ collider. We find no significant signal and
set an upper limit of 
$\mathcal{B} (B^+ \to D^{*+} \pi^0) < 3.6 \times 10^{-6}$  at the $90\%$
confidence level. This limit can be used to constrain the ratio between
suppressed and favored $B \to D^{*} \pi$ decay amplitudes, $r < 0.051$, 
at the $90\%$ confidence level.
\end{abstract}

\pacs{13.25.Hw, 11.30.Er, 12.15.Hh, 14.40.Nd}

\maketitle

{\renewcommand{\thefootnote}{\fnsymbol{footnote}}}
\setcounter{footnote}{0}

In the Standard Model, $CP$ violation arises from a complex phase in the
Cabibbo-Kobayashi-Maskawa (CKM) quark mixing matrix~\cite{Cabibbo,
KM}. Precise measurements of CKM matrix parameters are therefore of
fundamental importance for the description of the weak interaction of
quarks and the investigation for the new sources of $CP$
violation. Measurements of the time-dependent decay rates of 
$B^0(\overline{B}{}^0) \to D^{*\mp}\pi^{\pm}$ provide a theoretically
clean method for extracting $\sin(2\phi_1+\phi_3)$~\cite{sin2phi1phi3},
where $\phi_1$ and $\phi_3$ are the interior angles of the CKM
triangle~\cite{def_angle}.
The $CP$ violation parameters $S^\pm$ are given by~\cite{fleischer}
\begin{eqnarray}
S^\pm = \frac{2(-1)^L\, r\, \mathrm{sin}(2\phi_1 + \phi_3 \pm
 \delta)}{1+r^2},
\end{eqnarray}
where $r$ is the ratio of the amplitudes of the doubly Cabibbo
suppressed decay (DCSD), $B^0 \rightarrow D^{*+}\pi^-$ to the Cabibbo
favored decay (CFD), $B^0 \rightarrow D^{*-}\pi^+$ (Fig. 1), $L$ denotes
the angular momentum of the final state, and $\delta$ is the strong
phase difference between DCSD and CFD. It is difficult to determine $r$
from $B^0$ decays because the DCSD amplitude is small compared to the
contribution from mixing followed by CFD, 
$B^0 \to \overline{B}{}^{0} \to D^{*+} \pi^-$. 

Using available branching fraction measurements, $r$ can be 
expressed as
\begin{eqnarray}
r = \tan \theta_c \frac{f_{D^{*}}}{f_{D_{s}^{*}}}
\sqrt{\frac{\mathcal{B}(B^0 \to D_{s}^{*+} \pi^-)}{\mathcal{B}(B^0 \to
D^{*-} \pi^+)}},
\label{eq:r_ds}
\end{eqnarray}
where $\theta_c$ is the Cabibbo angle, and the decay constants
$f_{D^{*}}$ and $f_{D_{s}^{*}}$ are available from lattice QCD
calculations. However, the assumption of SU(3) symmetry and additional
$W$-exchange contributions result in an uncertainty of about $30\%$ on
$r$. In order to avoid this uncertainty, one can instead use the isospin
relation,
\begin{eqnarray}
r = \sqrt{ \frac{\tau_{B^0}}{\tau_{B^+}} \frac{2 \mathcal{B}(B^+ \to
 D^{*+} \pi^0)}{\mathcal{B}(B^0 \to D^{*-} \pi^+)}},
\label{eq:isospin}
\end{eqnarray}
where $\tau_{B^+}/\tau_{B^0} = 1.071 \pm 0.009$ and 
$\mathcal{B}( B^0 \to D^{*-}\pi^{+} ) = (2.76\pm0.21) \times 10^{-3}$~\cite{PDG}.
We naively estimate 
$\mathcal{B}( B^+ \to D^{*+}\pi^{0} ) = 5.9 \times 10^{-7}$,
taking into account the $r$ factor of $0.02$ calculated from 
Eq.~(\ref{eq:r_ds})~\cite{sin2phi1_phi3_paper_belle}. 
The previous search gives an upper limit of
$\mathcal{B}( B^+ \to D^{*+}\pi^{0} ) < 1.7 \times 10^{-4}$ at the
$90\%$ confidence level~\cite{cleo_search}.
\begin{center}
\begin{figure}[htbp]
\subfigure[\ $B^0 \to D^{*-} \pi^+$]
{\includegraphics[width=0.235\textwidth,bb=110 310 435 490,clip]{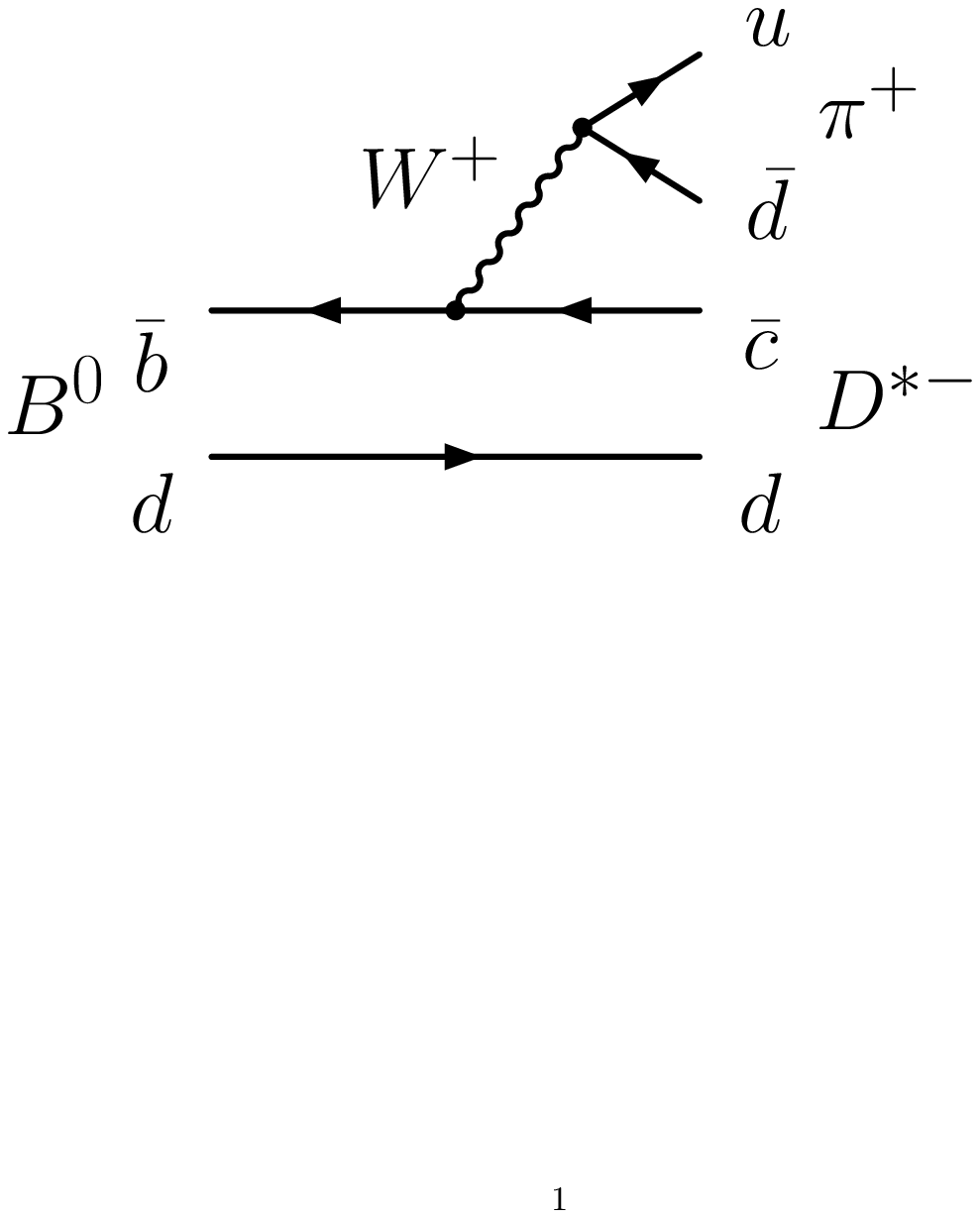}}
\subfigure[\ $B^{+(0)} \to D^{*+} \pi^{0(-)}$]
{\includegraphics[width=0.235\textwidth,bb=110 310 435 490,clip]{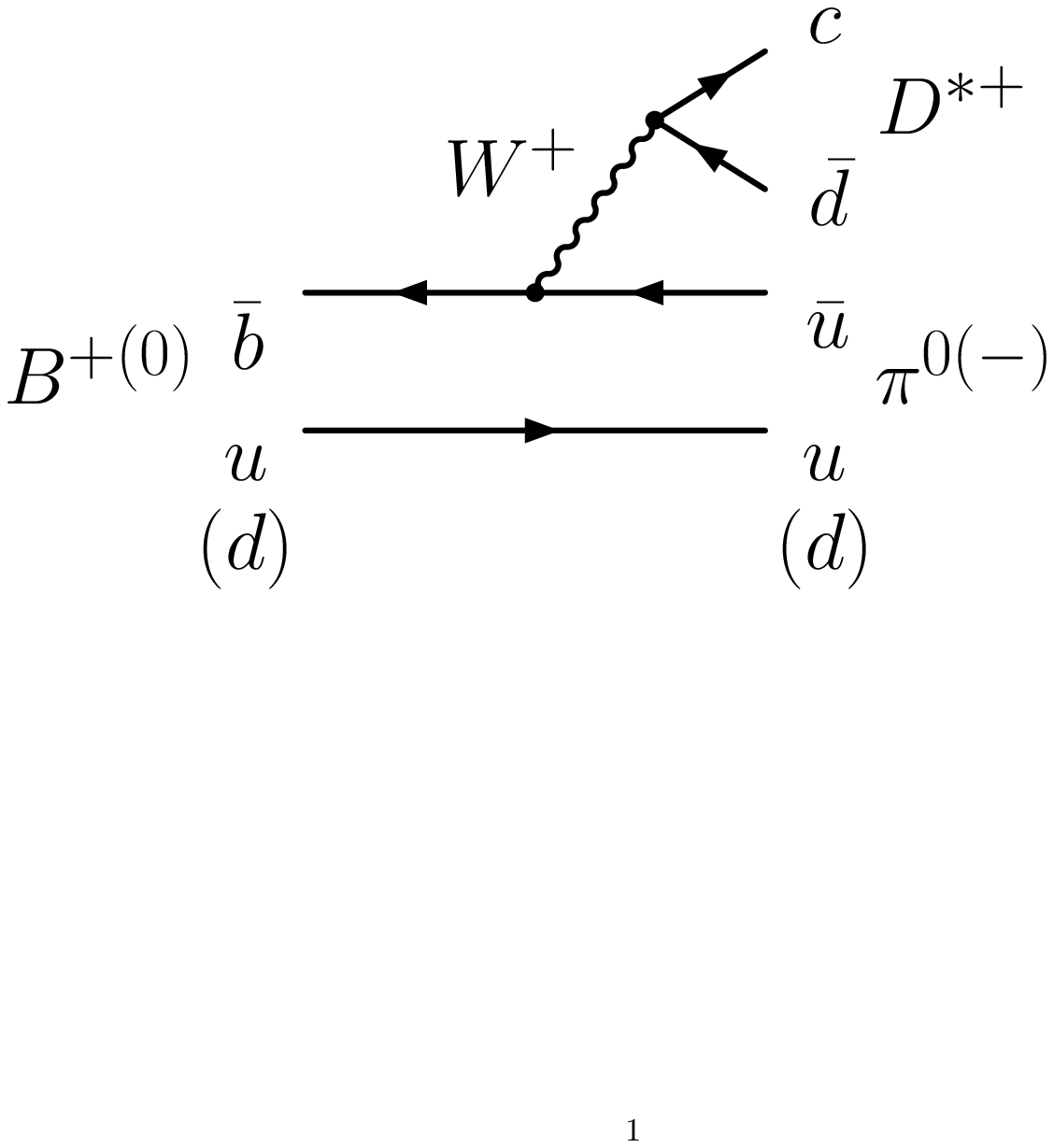}}
\caption{ 
Feynman tree diagrams for (a) CFD $B^0 \to D^{*-}\pi^{+}$ 
with the CKM coupling $V_{cb}^{*}V_{ud}$, and (b) DCSD 
$B^{+(0)} \to D^{*+}\pi^{0(-)}$ with the coupling 
$V_{ub}^{*}V_{cd}$.} 
\label{fig:cfd_dcsd}
\end{figure}
\end{center}

In this paper, we report on a search for \md \ based on a data
sample of $605 \ {\rm fb}^{-1}$ corresponding to
$(657 \pm 9) \times 10^6\, B\overline{B}$ events, collected  with the
Belle detector~\cite{Belle} at the KEKB asymmetric-energy $e^+e^-$
collider~\cite{KEKB} operating at the $\Upsilon(4S)$ resonance.

The Belle detector is a large-solid-angle magnetic spectrometer that
consists of a silicon vertex detector, a 50-layer central drift chamber
(CDC), an array of aerogel threshold Cherenkov counters (ACC), a
barrel-like arrangement of time-of-flight scintillation counters (TOF), 
and an electromagnetic calorimeter comprised of CsI(Tl) crystals located
inside a superconducting solenoid coil that provides a magnetic field of
1.5 T.  An iron flux-return located outside of the coil is instrumented
to detect $K_L^0$ mesons and to identify muons.  

To search for $B^+ \to D^{*+}\pi^{0}$,  we reconstruct $D^{*+}$
candidates by pairing a low momentum charged pion ($\pi^+_{\rm slow}$)
and a $D^0$, which is reconstructed through its decays to 
$K^-\pi^+$, $K^-\pi^+\pi^0$, $K^-\pi^+\pi^-\pi^+$, and
$K_S^0\pi^+\pi^-$.
Inclusion of charge conjugate modes is implied throughout this paper.

For charged kaon and pion candidates except pions from $K_S^0$'s,
we require tracks to have a distance of closest approach to the
interaction point within 5 cm along the $z$-axis (anti-parallel to the
positron beam direction) and within 2 cm in a plane perpendicular to the
$z$-axis. Particle identification (PID) is based on the likelihoods
$\mathcal{R}(K/\pi) = \mathcal{L}_K/(\mathcal{L}_K +\mathcal{L}_{\pi})$,
where $\mathcal{L}_K$ ($\mathcal{L}_{\pi}$) is the likelihood of
kaons (pions) derived from the TOF, ACC, and $dE/dx$ measurements in the
CDC. The PID selections, which are $\mathcal{R}(K/\pi) > 0.3 \ (< 0.3)$
for kaons (pions) are applied to all charged particles except pions from
$K_S^0$'s. The PID efficiencies are $94 \%$ $(91\%)$ for kaons (pions),
while the probability of misidentifying a pion as a kaon (a kaon as a
pion) is $12 \%$ ($6 \%$).

Neutral pions are formed from photon pairs with an invariant mass
between $0.118 \GeVcc$ and $0.150 \GeVcc$, corresponding to $\pm 3$
standard deviations ($\sigma$). The photon momenta are then recalculated
with a $\pi^0$ mass constraint. We require the $\pi^0$ momentum to be
greater than $0.2\GeVc$ in the center-of-mass system (c.m.s.), and the
photon energy to be greater than $0.1\GeV$ in the laboratory frame.

$K_S^0$ candidates are reconstructed from pion pairs of
oppositely-charged tracks with an invariant mass between $0.485\GeVcc$
and $0.510\GeVcc$, corresponding to $\pm 5 \sigma$. Each candidate must
have a displaced vertex with a flight direction consistent with that of
a $K_S^0$ meson originating from the interaction point.
Mass- and vertex-constrained fits are applied to obtain the 4-momenta of
$K_S^0$ candidates.

For $D^0$ selection, the invariant mass of the daughter particles is
required to be within $3 \sigma$ from the nominal $D^0$ mass, where
$\sigma$ ($\sim5 \MeVcc $) depends on the decay mode. $D^{*+}$
candidates are required to have a mass difference 
$\Delta M = M_{D\pi}-M_{D}$ within $3 \sigma$ from the nominal mass
difference, where $\sigma$ ($\sim0.5 \MeVcc $) depends on the decay
mode. Mass- and vertex-constrained fits are applied to $D^0$ and
$D^{*+}$ candidates. 

We reconstruct a $B^+$ candidate from a $D^{*+}$ and a $\pi^0$ candidate. 
We identify $B$ decays based on requirements on the energy difference
$\dE \equiv \sum_i E_i - E_{\rm beam}$ and the beam-energy constrained
mass $\mbc \equiv \sqrt{E_{\rm beam}^2 - |\sum_i \overrightarrow{p_i}|^2 }$,
where $E_{\rm beam}$ is the beam energy, and $\overrightarrow{p_i}$ and
$E_i$ are the momenta and energies of the daughters of the reconstructed
$B$ meson candidate, all in the c.m.s. We select candidates in a fit
region defined as $|\dE| < 0.25 \GeV$ and 
$5.20 \GeVcc < \mbc < 5.29 \GeVcc$. The signal region is defined as 
$|\dE| < 0.1 \GeV$ and $5.27 \GeVcc < \mbc < 5.29 \GeVcc$.

To suppress the background from continuum 
($e^+e^- \to q\overline{q}, \ q=u,d,s,c$) events, we calculate modified
Fox-Wolfram moments~\cite{SFW2} and combine them into a Fisher
discriminant. We calculate a probability density function (PDF) for this
discriminant and multiply it by PDFs for $\cos\theta_{B}$, $\Delta z$,
and $\cos\theta_h$, where $\theta_{B}$ is the polar angle between the
$B$ direction and the beam direction in the c.m.s., $\Delta z$ is the
displacement along the beam axis between the signal $B$ vertex and that
of the other $B$, and $\theta_h$ is the angle between the 
$\pi^+_{\rm slow}$ direction and the opposite of the $B$ momentum
in the $D^{*+}$ frame. The PDFs for signal, generic $B$ events and
continuum are obtained from GEANT3-based~\cite{GEANT} Monte Carlo
(MC) simulation. These PDFs are combined into a signal (background)
likelihood variable $\mathcal{L}_{\rm sig(bkg)}$; we then impose
requirements on the likelihood ratio 
$\mathcal{R} \equiv \mathcal{L}_{\rm sig} / (\mathcal{L}_{\rm
sig}+\mathcal{L}_{\rm bkg})$. 
Additional background suppression is achieved through the use of a
$B$-flavor tagging algorithm~\cite{TaggingNIM}, which provides a
discrete variable indicating the flavor of the tagging $B$ meson and a
quality parameter $r_{\rm tag}$, with continuous values ranging from $0$
for no flavor information to unity for unambiguous flavor
assignment. The backgrounds from continuum and generic $B$ events are
reduced by applying a selection requirement on $\mathcal{R}$ for events
in each $r_{\rm tag}$ region that maximizes the value of 
$N_{\rm sig}/\sqrt{N_{\rm sig}+N_{\rm bkg}}$, where $N_{\rm sig}$ and
$N_{\rm bkg}$ denote the expected signal and background yields in the
signal region, based on MC simulation. This requirement eliminates
$99\%$ $(94\%)$ of the background from continuum ($B$ decays) in the
signal region, while retaining $35\%$ of the signal.

The fraction of events with more than one candidate is $3\%$. We select
the best $D^{*+}\pi^0$ candidate based on the value of 
$\chi^2_{\rm tot} = \chi^2_{M(D^0)}+\chi^2_{\Delta M}+\chi^2_{M(\pi^0)}$, 
where each $\chi^2$ is defined as the squared ratio of the deviation of
the measured parameter from the expected signal value and the
corresponding resolution. The reconstruction efficiency is determined
to be $0.56\%$, using the fitting procedure described below for the
signal MC samples. The branching fractions of $D^{*+}$ and $D^0$ are
included in the efficiency~\cite{PDG}.

After the selection criteria are applied, the dominant
background sources in the fit region are the continuum events and
\dstrho, while other $B$ decays such as 
$B^- \rightarrow D^0 \rho^-$ and $\overline{B}{}^0 \to D^{*0} \pi^0$ 
have smaller contributions. To obtain the signal yield, we perform  an
unbinned two-dimensional (2D) extended-maximum-likelihood fit to the
$\dE$-$\mbc$ distributions in the fit region. The likelihood function
consists of the following components: signal, continuum background
($q\overline{q}$), \dstrho, and other $B$ decays.

The likelihood function for the signal is defined separately for each of
the four $D^0$ decay modes and unified using the available branching
fractions of the $D^0$ subdecays~\cite{PDG}, while those for
$q\overline{q}$ and backgrounds from $B$ decays are defined as the sum
of four $D^0$ decay modes. Each $\dE$ and $\mbc$ shape for the signal is
modeled by the sum of a Gaussian and a bifurcated Gaussian with means
and widths fixed to the values obtained from MC simulation. The $\dE$
and $\mbc$ PDFs for $q\overline{q}$ are modeled by a linear function and
an ARGUS function~\cite{ARGUS}, respectively. The backgrounds from
\dstrho \ and other $B$ decays are modeled by the superposition of
Gaussian distributions constructed from unbinned MC events, where the
width of each Gaussian represents the smoothing parameter for the
event~\cite{KEYS}. The \dstrho \ background forms a large peak in the
region $\dE < -0.1 \GeV$ and $5.27 \GeVcc < \mbc < 5.29 \GeVcc$. 
The size and shape of the \dstrho \ component strongly depend on the
fraction of the longitudinal helicity component ($|H_0|$); we use 
$|H_0| = 0.941$ from Ref.~\cite{dstrho_polarz}.

The following parameters are allowed to vary: $q\overline{q}$ PDF
parameters and yields of signal, $q\overline{q}$ and 
\dstrho \ components. The yield of other $B$ decays is fixed to the
branching fractions in Ref.~\cite{PDG}. 

\begin{center}
\begin{figure}[htbp]
{\includegraphics[width=0.5\textwidth,bb=5 8 575 284,clip]{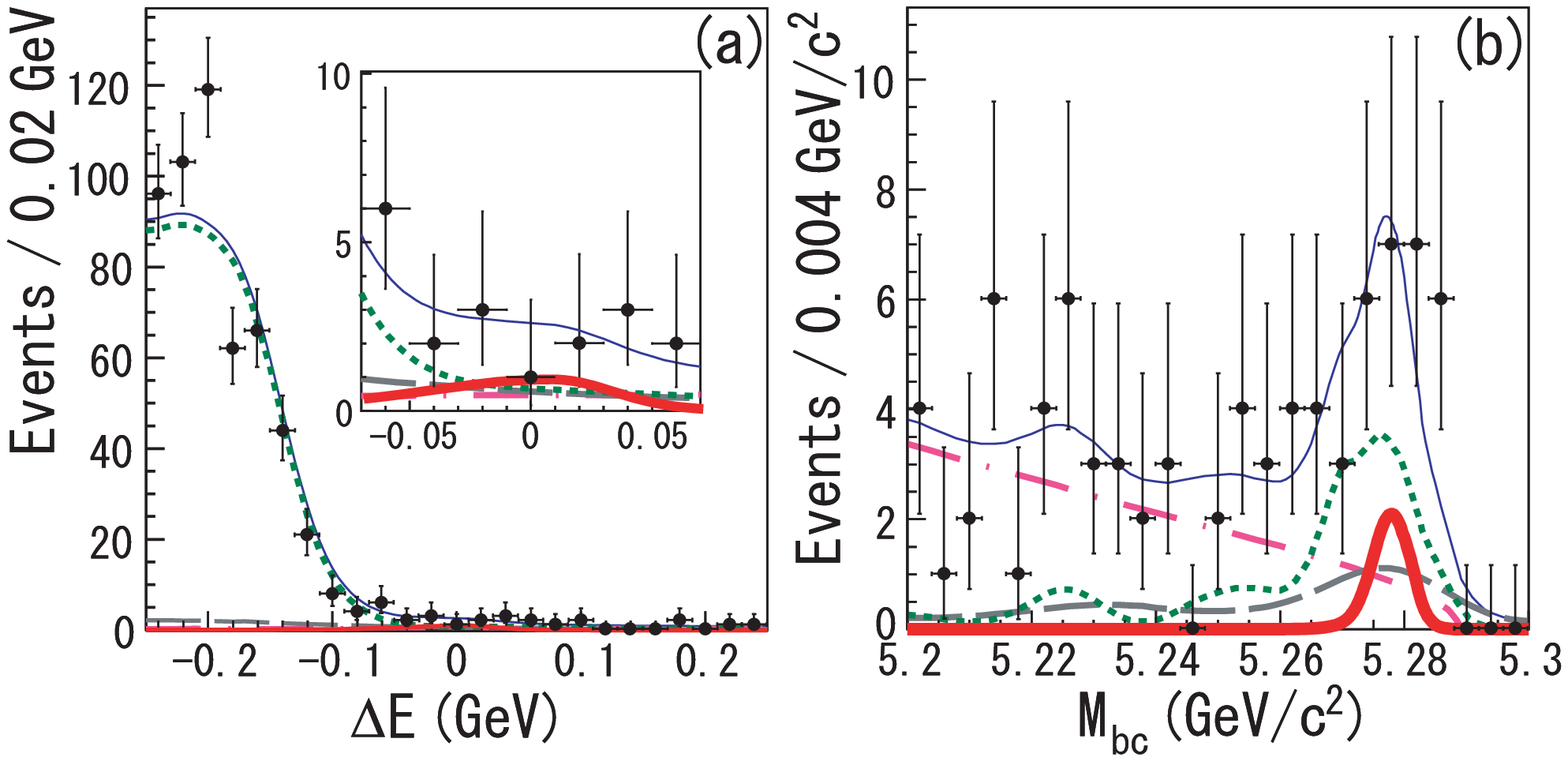}}
\caption{Projections of the unbinned two-dimensional likelihood fit to
 data in the region
 $|\dE|<0.25 \GeV $ and $5.20 \GeVcc <\mbc < 5.29 \GeVcc$. 
 (a) $\dE$ distribution for $5.27 \GeVcc < \mbc < 5.29 \GeVcc $ with a
 magnified view of $|\dE|<0.07 \GeV$ in the inset. (b) $\mbc$
 distribution for $|\dE| < 0.1 \GeV$. The points with error bars
 represent the data, while the curves represent the various components
 from the fit: signal (thick solid line), continuum (dash-dotted line), 
 $\overline{B}{}^0 \rightarrow D^{*+}\rho^-$ decay (dotted line), other
 $B$ decays (dashed line), and the sum of all components (thin solid
 line).}
\label{fig:fit_real}
\end{figure}
\end{center}

Figure \ref{fig:fit_real} shows the results of the fit to the data in
the fit region. The projections of the fitted $B$ signal in $\dE$
($\mbc$) in the $\mbc$ ($\dE$) signal region are shown. 
We obtain $4.5_{-3.4}^{+4.1}$ \md \ signal candidates in the
signal region (statistical error only). The significance is
$1.4\sigma$, defined by 
$\sqrt{-2\ln(\mathcal{L}_0/\mathcal{L}_{\rm max})}$ where 
$\mathcal{L}_{\rm max}$ ($\mathcal{L}_0$) is the likelihood value at the
maximum (with the signal fixed to zero). The likelihood function is
convolved with an asymmetric Gaussian distribution that represents the
systematic error.

The systematic error components proportional to the signal yield are
determined as follows. We estimate the systematic error from the
$\mathcal{R}$ requirement by applying the $\mathcal{R}$ requirement to
data and MC events using a $B^- \rightarrow D^0 \rho^-$ control sample. 
The systematic error on the $\Delta M$ requirement is estimated by
applying the $\Delta M$ requirement to 
$\overline{B}{}^0 \rightarrow D^{*+}\pi^-$ data and \md \ MC samples. 
The systematic error on the secondary branching fraction is calculated
from errors given in Ref.~\cite{PDG}. The systematic error due to the
charged-track reconstruction efficiency is estimated to be $1.0\%$
($1.6\%$) per charged kaon (pion) using partially reconstructed $D^{*+}$
events. The systematic error due to $\mathcal{R}(K/\pi)$ selection has a
relative uncertainty of $0.8\%$ ($1.4\%$) per charged kaon (pion),
determined from $D^{*+} \to D^0 \pi^+$, $D^0 \to K^- \pi^+$ decays. 
The $\pi^0$ reconstruction is verified by comparing the ratio of 
$D^0 \to K^- \pi^+$ and $D^0 \to K^- \pi^+ \pi^0$ yields with 
the MC expectation; an uncertainty of $3.0\%$ per particle is
assigned. The $K_S^0$ reconstruction is verified by comparing the ratio
of $D^+ \to K_S^0 \pi^+$ and $D^+ \to K^- \pi^+ \pi^+$ yields with 
the MC expectation; an uncertainty of $4.9\%$ is assigned. The
systematic error due to the signal MC statistics is $0.5\%$ and the
error due to the uncertainty in the total number of $B\overline{B}$
pairs is $1.4\%$. The systematic error components proportional to the
signal yield are summarized in Table~\ref{tab:syst_1}.

The systematic errors on the yield extraction are estimated as follows.
We estimate the uncertainty of $|H_0|$ of \dstrho \ by varying $|H_0|$
by $\pm1\sigma$, where the error of $|H_0|$ is taken from
Ref.~\cite{dstrho_polarz}. Possible $\dE$ shifts between data and MC
simulation for the \dstrho \ background are evaluated by measuring the
$\dE$ shift of the $B^- \rightarrow D^{*0}\rho^-$ background component
using a $\overline{B}{}^0 \rightarrow D^{*0}\pi^0$ control sample. To
obtain the systematic error on the background fraction of other $B$
decays, we vary the normalizations of the individual sources by
$\pm1\sigma$, where the values are taken from Ref.~\cite{PDG}. The
normalization of other background components are varied by
$\pm50\%$. The systematic error due to the uncertainty in the shape of
the $B$ background PDF is determined by varying the Gaussian smoothing
width by factors of two and one half from its nominal
value. Uncertainties from the two-dimensional correlation in the signal
and the $q\overline{q}$ components are estimated by applying 2D
background PDFs to the signal and the $q\overline{q}$ shapes. The effect
of a possible bias in the fitting procedure is estimated by a toy MC
study. The systematic errors on the yield extraction in the signal
region are summarized in Table~\ref{tab:syst_2}.

\begin{table}[htbp]
\caption{Systematic errors for $\mathcal{B}$(\md), proportional to the
 signal yield.}
\label{tab:syst_1}
\begin{tabular}{lc}
\hline \hline
Source     & Systematic error ($\%$) \\
           &  $\pm \sigma$  \\
\hline
$\mathcal{R}$ requirement        &  $3.0$ \\
$\Delta M$ requirement           &  $3.3$ \\
Secondary branching fractions    &  $3.3$ \\
Track finding efficiency         &  $5.1$ \\
Particle identification          &  $4.4$ \\
$\pi^0$ reconstruction           &  $4.1$ \\
$K_S^0$ reconstruction           &  $0.3$ \\
MC statistics                    &  $0.5$ \\
Number of $B\overline{B}$ pairs  &  $1.4$ \\
\hline
Quadratic sum                 &  $9.8$ \\
\hline
\end{tabular}
\end{table}
\begin{table}[htbp]
\caption{Systematic errors for $\mathcal{B}$(\md), related to the yield
 extraction in the signal region.}
\label{tab:syst_2}
\begin{tabular}{p{11em}p{5em}p{5em}}
\hline \hline
Source     & \multicolumn{2}{c}{Systematic error} \\ 
           & \multicolumn{2}{c}{(number of events)} \\
           & \hfil $+\sigma$ \hfil  & \hfil  $-\sigma$ \hfil  \\
\hline
$|H_0|$ of \dstrho                   & \hfil $0.7$  \hfil
                                     & \hfil $-1.9$ \hfil \\
$\dE$ shift of \dstrho               & \hfil $0.0$  \hfil 
                                     & \hfil $-0.6$ \hfil \\
Fraction of backgrounds              & \hfil $0.8$  \hfil
                                     & \hfil $-0.4$ \hfil \\
Gaussian width of                    & \hfil $0.5$  \hfil
                                     & \hfil $-2.0$ \hfil \\
\ \ \ \ background PDF \\
2D correlation for $q\overline{q}$   & \hfil $0.0$  \hfil 
                                     & \hfil $-1.3$ \hfil \\
\ \ \ \ and \md \\
Fit bias                             & \hfil $0.0$  \hfil 
                                     & \hfil $-0.5$ \hfil \\ 
\hline
Quadratic sum                        & \hfil $1.2$  \hfil
                                     & \hfil $-3.2$ \hfil \\
\hline
\end{tabular}
\end{table}

We then obtain the branching fraction of \md \ to be
$\mathcal{B}(B^+ \to D^{*+} \pi^0) = [ 1.2_{-0.9}^{+1.1}({\rm stat})_{-0.9}^{+0.3}({\rm syst}) ] \times 10^{-6}$.

The likelihood distribution (${\cal L}$), which is convolved with the
systematic error, is used to obtain the upper limit on the branching
fraction. We calculate the $90\%$ confidence level (C.L.) upper limit
(UL) using the relation 
$\int_0^{\rm UL}{\cal L} d\mathcal{B} / \int_0^{\infty}{\cal L} d\mathcal{B} = 0.9 $ 
to be 
\begin{eqnarray}
\mathcal{B}(B^+ \to D^{*+} \pi^0) < 3.6 \times 10^{-6}.
\end{eqnarray}
\noindent
The obtained upper limit is consistent with the naive estimate, 
$5.9 \times 10^{-7}$ discussed above. This result can be used to obtain
an upper limit on the ratio of magnitudes of DCSD and CFD in $D^*\pi$
decay,
\begin{eqnarray}
r < 0.051 \ \ (90 \%\ {\rm C.L.}).
\end{eqnarray}

To summarize, a search for the doubly Cabibbo suppressed decay \md \ in
a data sample of $605 \ {\rm fb}^{-1}$ yields an upper limit of 
$\mathcal{B}(B^+ \to D^{*+} \pi^0) < 3.6 \times 10^{-6}$
at the $90\%$ confidence level. This limit can be used to constrain the
ratio between suppressed and favored $B \to D^{*} \pi$ decay amplitudes,
$r < 0.051$, at the 90\% confidence level.

We thank the KEKB group for excellent operation of the accelerator, the
KEK cryogenics group for efficient solenoid operations, and the KEK
computer group and the NII for valuable computing and
SINET3 network support.  We acknowledge support from MEXT and JSPS (Japan);
ARC and DEST (Australia); NSFC (China); DST (India); MOEHRD, KOSEF and
KRF (Korea); KBN (Poland); MES and RFAAE (Russia); ARRS (Slovenia); SNSF
(Switzerland); NSC and MOE (Taiwan); and DOE (USA).


\begin{thebibliography}{99}

\bibitem{Cabibbo}
N.~Cabibbo, Phys. Rev. Lett. {\bf 10}, 531 (1963).

\bibitem{KM}
M.~Kobayashi and T.~Maskawa, Prog. Theor. Phys. {\bf 49}, 652 (1973).

\bibitem{sin2phi1phi3} 
I.~Dunietz and R.~G.~Sachs, Phys. Rev. D {\bf 37}, 3186 (1988); Erratum
	{\it ibid.} {\bf 39}, 3515 (1989); I. Dunietz, Phys. Lett. B
	{\bf 427}, 179 (1998). 

\bibitem{def_angle} 
The angles $\phi_1$ and $\phi_3$ are also sometimes known as $\beta$ and
	$\gamma$, respectively.

\bibitem{fleischer} R.~Fleischer, Nucl. Phys. {\bf B671}, 459 (2003).

\bibitem{PDG}
Y.-M.~Yao {\it et al.} (Particle Data Group), J. Phys. G {\bf 33}, 1 (2006).

\bibitem{sin2phi1_phi3_paper_belle} 
F.~J.~Ronga and T.~R.~Sarangi {\it et al.} (Belle Collaboration),
	Phys. Rev. D {\bf 73}, 092003 (2006).

\bibitem{cleo_search} 

G.~Brandenburg {\it et al.} (CLEO Collaboration), Phys. Rev. Lett. {\bf 80},
	2762 (1998).

\bibitem{Belle}
A.~Abashian {\it et al.} (Belle Collaboration), Nucl. Instrum. Methods
	Phys. Res., Sect. A {\bf 479}, 117 (2002).

\bibitem{KEKB}
S.~Kurokawa and E.~Kikutani, Nucl. Instrum. Methods
	Phys. Res., Sect. A {\bf 499}, 1 (2003), and other papers
	included in this volume.

\bibitem{SFW2}
G.~C.~Fox and S.~Wolfram, Phys. Rev. Lett. {\bf 41}, 1581 (1978). The
	modified moments used in this paper are described in, S.~H.~Lee
	{\it et al.} (Belle Collaboration), Phys. Rev. Lett. {\bf 91},
	261801 (2003).

\bibitem{GEANT}
R.~Brun {\it et al.}, CERN Report No. DD/EE/84-1 (1984), {\tiny GEANT}
	3.21.

\bibitem{TaggingNIM}
H.~Kakuno {\it et al.}, Nucl. Instrum. Methods
	Phys. Res., Sect. A {\bf 533}, 516 (2004). 

\bibitem{ARGUS}
H.~Albrecht {\it et al.} (ARGUS Collaboration), Phys. Lett. B {\bf 241},
	278 (1990).

\bibitem{KEYS} 
K.~Cranmer, Comput. Phys. Commun. {\bf 136}, 198 (2001).

\bibitem{dstrho_polarz} 
S.~E.~Csorna {\it et al.} (CLEO Collaboration), Phys. Rev. D {\bf 67},
	112002 (2003).

\end{thebibliography}
\end{document}